\documentclass[%
 reprint,
 amsmath,amssymb,
 aps,
]{revtex4-2}
\usepackage{amssymb}
\usepackage{amsthm}
\usepackage{graphicx}
\usepackage{dcolumn}
\usepackage{bm}

\begin{document}

\title{Statistical mechanics for  non-Hermitian quantum systems }
\author{Kui Cao}
\affiliation{Center for Advanced Quantum Studies, Department of Physics, Beijing Normal
University, Beijing 100875, China}
 
\author{Su-Peng Kou}
\thanks{Corresponding author}
\email{spkou@bnu.edu.cn}
\affiliation{Center for Advanced Quantum Studies, Department of Physics, Beijing Normal
University, Beijing 100875, China}

\begin{abstract}
We present a systematic study of  statistical mechanics for  non-Hermitian quantum systems. Our work reveals that the stability of a non-Hermitian system necessitates the existence of a single path-dependent conserved quantity, which, in conjunction with the system's Hamiltonian, dictates the equilibrium state. By elucidating the relationship between the Hamiltonian and the supported conserved quantity, we propose criteria for discerning equilibrium states with finite relaxation times. Although our findings indicate that only non-Hermitian systems with real energy spectrum precisely possess such conserved quantities, we also demonstrate that an effective conserved quantity can manifest in certain systems with complex energy spectra. The effective conserved quantity, alongside the effective transitions within their associated subspace, collectively determines the system's equilibrium state. Our results provide valuable insights into non-Hermitian systems across more  realistic contexts and hold potential for applications in a diverse range of physical systems. 

\end{abstract}

\maketitle

\section{introduction}  
In recent years, non-Hermitian quantum systems have garnered significant interest due to their unique and intriguing properties. These systems exhibit a rich variety of phenomena, including exceptional points (EPs) \cite{Bender98,Bender02,Bender07}, non-Hermitian skin effect \cite{Yao2018,Yao20182,Kunst2018,Yin2018,KawabataUeda2018,Xiong2018,Torres2018,Ghatak2019,Lee2019,Yokomizo2019,KZhang2020,Slager2020,YYi,SMu2020,Kazuki2021,Okuma2021,Roccati,HLiu2022,CF2022}, and the breakdown of bulk-boundary correspondence \cite{Yao2018,Xiong2018,Yao20182,Kunst2018,Louc}. Most of the research focus has been on the ground state properties of these systems. However, in realistic scenarios, these systems are inevitably coupled with environment, leading to a non-coherent superposition of excited and ground states. Such situations are generally treated using statistical mechanics. To develop a more comprehensive understanding of non-Hermitian quantum states, it is essential to consider the statistical mechanics of  non-Hermitian systems. 

In the context of statistical mechanics, conserved quantities play a pivotal role in determining the equilibrium properties of a system. The probability of a system's microstate occurring is determined by the associated conserved quantity (typically energy). The vast majority of research involving non-Hermitian quantum systems at finite temperatures has treated  the statistical mechanics of non-Hermitian quantum systems as a direct extension of Hermitian quantum systems---the equilibrium state density matrix is determined by the system's Hamiltonian, which encodes the energy of each state \cite{Herviou2019,Quthe,LS2015,CZ2018,YTB2020,YNKTU2022,ZDBSN2022,Chang2020,Zhang2020,Chen2022}. However, non-Hermitian quantum systems, as a distinct class of open systems, do not adhere to strict energy conservation despite having a time-translation invariant  Hamiltonian. This deviation from Hermitian systems implies that the original statistical mechanics framework may no longer be applicable to non-Hermitian systems. In fact, the lack of strict energy conservation may even render the existence of a stable state questionable \cite{Hamazaki2020,Cipolloni2023,Cipolloni20232}.  

In this paper, we conduct  a systematic study of quantum non-Hermitian systems coupled to thermal bath, investigating their thermodynamic properties and stability conditions. We have found that the stability of non-Hermitian systems necessitates  the exist of path-dependent conserved quantity. The path-independent part of the conserved quantity, together with the system's Hamiltonian, determines the equilibrium state of the non-Hermitian system. As this conserved quantity constrains the thermalization path of the system, for certain conserved quantities, the thermalization path becomes, in a sense, extremely long, resulting in divergent relaxation times and making the equilibrium state challenging to observe. In order to exclude these difficult-to-observe equilibrium states, we have established a relationship between the non-Hermitian system's Hamiltonian and the supported conserved quantity, which aids in determining which equilibrium states have finite relaxation times. Furthermore, although we discovered that only non-Hermitian systems with real energy spectra precisely possess such conserved quantities, we have found that in more general cases with complex energy spectra, even though the system lacks an exact conserved quantity, when the thermal bath and system are weakly coupled, an effective conserved quantity rapidly emerges within this system. This conserved quantity, in conjunction with the effective transitions occurring within its associated subspace, jointly determines the system's equilibrium state.

This paper is organized in the following way. In Section II, we introduce the system under investigation and outline some basic assumptions. In Section III and Section IV, we develop the statistical mechanics theory for quasi-Hermitian systems and more general non-Hermitian systems, providing criteria to determine the stability of such systems and identifying their equilibrium states when stability is present. In Section V, we showcase the application of our theory using a $\mathcal{PT}$-symmetric two-level model as an example. Finally, in Section VI, we conclude our results and discuss the implications of our findings for the understanding of non-Hermitian  systems in various contexts. 

\section{The Framework}

The traditional research areas in statistical mechanics can be divided into two distinct categories: (i) systems weakly coupled  with a large thermal bath, and (ii) isolated systems. In this study, our focus lies on the former, specifically investigating the non-Hermitian system in contact with a \textit{Hermitian} thermal bath. This choice is motivated by the fact that, in most experimental scenarios, non-Hermiticity is prepared solely within a small system, while both the environment and the coupling between the environment and the system remain Hermitian. Consequently, our investigation focuses on  systems weakly interacting with Hermitian thermal baths, as this approach is better aligned with the practical aspects of experimental setups. On the other hand, isolated  non-Hermitian systems may lack the capacity for self-thermalization, rendering the establishment of a statistical theory for isolated non-Hermitian systems unattainable \cite{Hamazaki2020,Cipolloni2023,Cipolloni20232}.

The time evolution equation of the density matrix for a non-Hermitian system is given by \cite{BrodyGraefe2012,Kawabata2017,QWG3}:
\begin{equation}
i\frac{d}{dt}\rho= \hat{H}\rho- \rho \hat{H}^{\dag} + [ \mathrm{tr} (\hat{H}^\dag-\hat{H}) \rho ] \rho,
\end{equation}
where $\rho$ represents the density matrix. The first two terms correspond to the standard Liouville evolution, while the third term ensures the normalization of the density matrix.  In this paper, for the sake of mathematical simplicity, we always assume that the dimension of the Hilbert space is finite. The Hamiltonian of the entire system, denoted by $\hat{H}_{tot}$, comprises three components:
\begin{equation}
\hat{H}_{tot}=\hat{H}_{\mathrm{NH}}\otimes \hat{I}_B+\hat{I}_S\otimes \hat{H}_{B}+\hat{H}_{BS},
\end{equation}
where $\hat{H}_{\mathrm{NH}}$ represents the non-Hermitian Hamiltonian of the system $S$, $\hat{H}_{B}$ corresponds to the Hamiltonian of the thermal bath, and $\hat{H}_{BS}$ denotes the coupling between the system and the thermal bath. It is important to note that, under this definition, an imaginary number can be added to the non-Hermitian Hamiltonian without affecting the time evolution equation of the density matrix. Therefore, in this paper, for simplicity, we always adjust the maximum imaginary part of the eigenvalue of the non-Hermitian system to zero. The coupling term, $\hat{H}_{BS}$, can be generally expressed as:
\begin{equation}
\hat{H}_{BS} = \sum_{a} \lambda_{a} \hat{C}_{a} \otimes \hat{B}_{a} + H.c.,
\end{equation}
where $\lambda_{a}$ is a real number quantifying the strength of coupling
between the system and the thermal bath, $\hat{C}_{a}$ is the operator acting on the system, and $\hat{B}_{a}$ is the operator acting on the thermal bath, $a=1,2,3,...,n$. There are often multiple expressions for the same coupling. We choose the operators $\hat{C}_{a}, \hat{C}^{\dag}_{a}$ and $\hat{B}_{a}, \hat{B}^{\dag}_{a}, \hat{I}_B$ to be linearly independent. We set the operator acting on the thermal bath and $\hat{I}_B$ to be linearly independent to prevent the Hamiltonian of the system from being written in the form of similar coupling, making the system's Hamiltonian and coupling can not distinguish clearly.

 \begin{figure}[ptb]
\centering
\includegraphics[width=9cm]{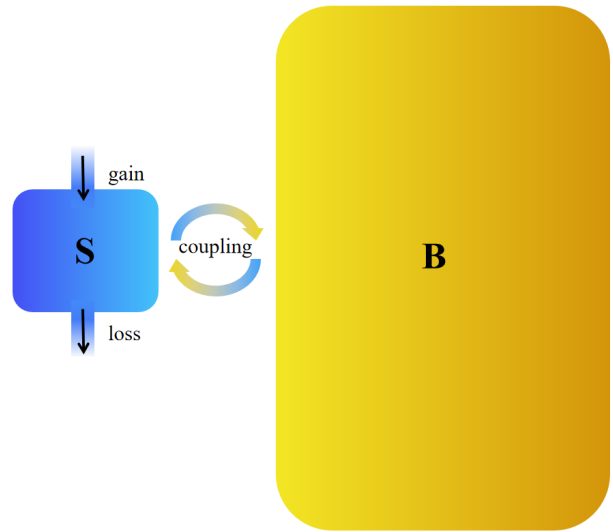}
\caption{Illustration for a thermal non-Hermitian system: $S$ represents the non-Hermitian system, while $B$ denotes a Hermitian thermal bath at centain temperature $T$. The coupling between the non-Hermitian system   and the Hermitian thermal bath is also Hermitian.}
\end{figure}

We have assumed that both $\hat{H}_{B}$ and $\hat{H}_{BS}$ are Hermitian. Owing to the weak coupling condition, we consider $  \lambda_{a}  \ll 1$. This system is referred to as a \textit{thermal non-Hermitian system}. See illustration in Fig. 1. The expectation value of a physical quantity for a mixed state is defined in the standard way. That is, the expectation value of the physical quantity $\hat{O}$ is given by:
\begin{equation}
\langle \hat{O}\rangle = \frac{1}{\mathrm{tr}\rho}\ \mathrm{tr}[\hat{O}\rho].
\end{equation}
In developing the statistical mechanics for non-Hermitian systems, the concept of temperature is indispensable. We define the temperature of  non-Hermitian system as follows:

 \textit{Definition 1:} The temperature of the non-Hermitian system is defined as the temperature of the Hermitian bath if the corresponding thermal non-Hermitian system can reach an equilibrium state. 

The equilibrium state is defined as the steady state for thermalizable thermal non-Hermitian systems. We define the steady state in a common way, that is, the density matrix when time tends to infinity (if the limit exists). The definition of the ``thermalizable" will be provided later. It is worth noting that this definition reduces to the conventional temperature definition in the case of a Hermitian system, ensuring consistency and compatibility with well-established thermodynamic concepts.

\section{Quantum Statistical Mechanics for Quasi-Hermitian Systems}

In this section, we focus on  the thermal quasi-Hermitian systems, which involve a quasi-Hermitian system coupled with a Hermitian thermal bath. A quasi-Hermitian system is characterized as a real-spectrum system (In this paper, we consistently avoid EPs and treat them as a limiting case for systems without EPs). The $\mathcal{PT}$-symmetric system at $\mathcal{PT}$-symmetric phase serves as a prime example of a quasi-Hermitian system. While seemingly simple, the equilibrium state of a quasi-Hermitian system lays the ground for comprehending the equilibrium state of a general non-Hermitian system. We will later illustrate that the equilibrium problems of non-Hermitian systems can be reduced to those of quasi-Hermitian systems.

\subsection{Stability of quasi-Hermitian systems}

A quasi-Hermitian system represents a non-Hermitian configuration characterized by a delicate balance between gain and loss. This delicate balance renders the system susceptible to perturbations that may induce exponential amplification of specific state probabilities over time, significantly affecting the system's steady state. This phenomenon exposes the inherent instability of such systems.

To preserve system stability, it is important to maintain a purely real energy spectrum, thereby safeguarding the overall system balance. In the case of thermal quasi-Hermitian systems, individual subsystems and their corresponding thermal baths exhibit real spectrum. Nevertheless, the introduction of coupling often disturbs the real spectrum of the complete system. Consequently, the steady state heavily relies on the intricacies of the coupling between the bath and the system, as well as on the subtle details of the bath Hamiltonian. A discussion based on the master equation is provided in Appendix B for further technical insights. We refer to this phenomenon, in which all components are Hermitian or quasi-Hermitian but lose the quasi-Hermitian property upon combination, as \textit{quasi-Hermiticity breaking}. A thermal quasi-Hermitian system without quasi-Hermiticity breaking is   stable.

The real eigenvalues condition implies that a system can be transformed, at a minimum, into a Hermitian matrix in the form of $diag(\lambda_{1},\lambda_{2},\lambda_{3}...)$ via a similarity transformation, with $\lambda_{i}$ representing the eigenvalues. This leads to the following assertion, which emphasizes a   attribute concerning the absence of  quasi-Hermiticity breaking: For systems devoid of quasi-Hermiticity breaking, a similarity transformation must exist that transforms the total Hamiltonian $\hat{H}_{tot}$ into a Hermitian form.

Moreover, due to the Hermitian nature of the thermal bath, the similarity transformation takes the form $\mathcal{\hat{S}}_{\mathrm{tot}}=\mathcal{\hat{S}}\otimes \hat{I}_B$. The right side of the direct product can be any unitary transformation, which is tantamount to a basis transformation and does not influence the physical interpretation. We do not differentiate this transformation from the identity transformation. The operator $\mathcal{\hat{S}}$ can transform the non-Hermitian operator $\hat{H}_{\mathrm{NH}}$ into a Hermitian operator $\hat{H}_{0}$, i.e., $\mathcal{\hat{S}}^{-1}\hat{H}_{\mathrm{NH}}\mathcal{\hat{S}}=\hat{H}_{0}$. The remaining question is whether the coupling $\hat{H}_{BS}$ can retain its Hermitian property under a similarity transformation. 

\textit{Lemma 1}: A thermal non-Hermitian system is a thermal quasi-Hermitian system without quasi-Hermiticity breaking if and only if there exists a positive definite Hermitian operator $\mathcal{T}_c$ in the non-Hermitian system such that the following two conditions are satisfied:

(1) Symmetric condition: the coupling operator $\hat{C}_{a}$ satisfies $[\hat{C}_{a},\mathcal{T}_c]=0$ for all $a$. Alternatively, in physical terms, the \textit{coupling} needs to have symmetry  $\mathcal{T}_c$.

(2) Conjugacy relation: $\hat{H}_{\mathrm{NH}}\mathcal{T}_c-\mathcal{T}_c\hat{H}_{\mathrm{NH}}^{\dagger}=0$.

\begin{proof}
On the one hand, we demonstrate that an operator satisfies conditions (1) and (2)   must exist for the thermal non-Hermitian system without quasi-Hermiticity breaking. We have shown that the coupling after the similarity transformation needs to be a Hermitian operator. This requires
\begin{align}
& \sum_a \lambda_{a}[\mathcal{\hat{S}}^{-1}\hat{C}_{a}\mathcal{\hat{S}}\otimes \hat{B}_{a}+\mathcal{\hat{S}}^{-1}\hat{C}_{a}^{\dag}\mathcal{\hat{S}}\otimes \hat{B}_{a}^{\dag}] \notag  \\
& =\sum_a \lambda_{a}[(\mathcal{\hat{S}}^{-1}\hat{C}_{a}^{\dag}\mathcal{\hat{S}})^{\dag}\otimes \hat{B}_{a}+(\mathcal{\hat{S}}^{-1}\hat{C}_{a}\mathcal{\hat{S}})^{\dag}\otimes \hat{B}_{a}^{\dag}].
\end{align}
Because   the difference operators are linear indepand in the right of the direct product,
we have $\mathcal{\hat{S}}^{-1}\hat{C}_{a}\mathcal{\hat{S}=}(\mathcal{\hat{S}%
}^{-1}\hat{C}_{a}^{\dag}\mathcal{\hat{S}})^{\dag}.$

Therefore, by using $\mathcal{\hat{S}}^{-1}\hat{C}_{a}\mathcal{\hat{S}%
=}(\mathcal{\hat{S}}^{-1}\hat{C}_{a}^{\dag}\mathcal{\hat{S}})^{\dag}$, we get\
\[
\mathcal{\hat{S}}(\mathcal{\hat{S}}^{-1}\hat{C}_{a}\mathcal{\hat{S})\mathcal{\hat{S}%
}^{\dag}=\hat{S}}(\mathcal{\hat{S}}^{-1}\hat{C}_{a}^{\dag}\mathcal{\hat{S}%
})^{\dag}\mathcal{\mathcal{\hat{S}}^{\dag}}
\]

or%
\begin{equation}
\hat{C}_{a}\mathcal{T}_c=\mathcal{T}_c\hat{C}_{a}^{{}}.
\end{equation}
  We find that $\hat{C}_{a}$ and $\mathcal{T}_c \equiv \mathcal{\hat{S}\mathcal{\hat{S}}^{\dag} }$ are commutative. It is evident that the operator $\mathcal{T}_c$   defined in this way is Hermitian, and its eigenvalues are equal to the squares of the singular values of the invertible operator $\mathcal{\hat{S}}$. Consequently, $\mathcal{T}_c$  is a positive definite Hermitian operator. 

It is also easy to verify that since $\mathcal{T}_c$ is related to the similarity transformation that changes $\hat{H}_{\mathrm{NH}}$ into a Hermitian form, $\hat{H}_{\mathrm{NH}}$ and $\mathcal{T}_c$ satisfy the relation $\hat{H}_{\mathrm{NH}}\mathcal{T}_c-\mathcal{T}_c\hat{H}_{\mathrm{NH}}^{\dagger}=0$. This is, in fact, another equivalent definition of quasi-Hermiticity for the Hamiltonian $\hat{H}_{\mathrm{NH}}$ \cite{Williams1969,Scholtz1992}.

On the other hand, if there is an operator $\mathcal{T}_c$ satisfying conditions (1) and (2), we can define $\mathcal{\hat{S}}=\sqrt{\mathcal{T}_c}$. Because $\mathcal{T}_c$ is a positive definite Hermitian operator, such an $\mathcal{\hat{S}}$ always exists, and we can verify that it transforms the system's Hamiltonian into a Hermitian operator which has real spectrum (it should be noted that the similarity transformation used to transpose the system's Hamiltonian into a Hermitian operator isn't unique. Alternative methodologies for constructing it exist. As an illustration, consider $\mathcal{\hat{S}}=\sqrt{\mathcal{T}_c} \ \hat{U}$, where $\hat{U}$ is defined as a unitary operator). Thus, we have proven the lemma 1.
\end{proof}
The  condition (1) (2) of Lemma 1 can be combined into a single condition, which is encapsulated by the following theorem:

\textit{Theorem 1 :} A thermal non-Hermitian system is a thermal quasi-Hermitian system without quasi-Hermiticity breaking if and only if there exists a conserved quantity in the form of $P_c(t) \mathcal{T}_c$ within the non-Hermitian system. Here $P_c(t) \equiv 
T_{t} \exp(2\int_{0}^{t}dt\left \langle \hat{\varUpsilon}\right \rangle
) $ \ is a path-dependence factor. Where
$T_{t}$ is the time order operator. $\left \langle \hat{\varUpsilon}\right \rangle \  $is
the expectation value of $\hat
{\varUpsilon}$ for non-Hermitian system's density matrix $\rho (t)$ at time $t$, and $\hat
{\varUpsilon}$ is defined as the non-Hermitian part of non-Hermitian Hamiltonian, i.e., $  \hat
{\varUpsilon} \equiv  \frac{1}{2i}(\hat{H}_{\mathrm{NH}} 
-\hat{H}_{\mathrm{NH}}^{\dagger}) $.  $\mathcal{T}_c$ is a  positive defined Hermitian operator in the non-Hermitian system.

\begin{proof}
Firstly, we prove that the thermal quasi-Hermitian system without  quasi-Hermiticity breaking has this conserved quantity. The condition for $P(t)_c\mathcal{T}_c$ to be a conserved quantity
operator is that $\frac{d}{dt} \mathrm{tr} [P_c(t)\mathcal{T}_c \rho(t)]=0$ holds for all initial density.  This requirement is equivalent to
\begin{equation}
 (\hat{H}_{\mathrm{NH}}\mathcal{T}_c-\mathcal{T}_c\hat{H}_{\mathrm{NH}%
}^{\dagger})+ [\mathcal{T}_c,\hat{H}_{BS}]=0.
\end{equation}
Applying Lemma 1, we get that both terms on the left-hand side of the equation are zero. Therefore, $P(t) \mathcal{T}_c$ is a conserved quantity.

Secondly, we prove that if there is such a conserved quantity, the thermal non-Hermitian system is a thermal quasi-Hermitian system without quasi-Hermiticity breaking. If there is such a conserved quantity, we require that
$  (\hat{H}_{\mathrm{NH}}\mathcal{T}_c-\mathcal{T}_c\hat{H}_{\mathrm{NH}%
}^{\dagger})+ [\mathcal{T}_c,\hat{H}_{BS}]=0$.
More detail,
\begin{align}
 &(\hat{H}_{\mathrm{NH}}\mathcal{T}_c-\mathcal{T}_c^{}\hat{H}_{\mathrm{NH}%
}^{\dagger})\otimes \hat{I}_B \notag  \\    &+ \sum_a \lambda_{a}([\mathcal{T}_c,\hat{C}_a
]\otimes \hat{B}_a +  [\mathcal{T}_c,\hat{C}^\dag_a]\otimes \hat{B}^\dag_a) =0,
\end{align}
because $\hat{B}_{a}, \hat{B}^{\dag}_{a}, \hat{I}_B$
are chosen to be linearly independent, we get the two conditions of Lemma 1. Therefore, we proved the theorem.
\end{proof}

We also have proved that the existence of conserved quantity is equivalent to the two conditions of Lemma 1. Therefore, technically, Lemma 1 can be used to specifically judge whether this conserved quantity exists in the system.
We call conserved quantities like $P_c(t) \mathcal{T}_c$ as path-dependent conserved quantities. In subsection B, we will demonstrate that $\mathcal{T}_c$, which is the path-independent part of the conserved quantity and quasi-Hermitian system Hamiltonians $\hat{H}_{\mathrm{NH}}$, collectively determines the equilibrium state of the system. It is important to note that the existence of such quasi-conserved quantities does not restrict   the final state  of the system for a given  initial state, but rather constrains the thermalization path. To some extent, certain paths are very long, implying that the equilibrium state have these conserved quantities does not easy to observe. These aspects will be discussed in subection C. The concept of path-dependent conserved quantities plays a pivotal role in understanding the intricate dynamics and phenomena in quasi-Hermitian  and  non-Hermitian  systems.

\subsection{Equilibrium state of stable quasi-Hermitian systems}
In this subsection, we discuss the equilibrium state of stable quasi-Hermitian systems. When the thermal bath is incorporated into the total Hamiltonian, the entire large system becomes an isolated system. In previous research, the \emph{Eigenstate Thermalization Hypothesis} (ETH) \cite{JoshuM,Srednicki1994,Deutsch1991,Rigol2008,Rigol2009,Biroli2010,Steinigeweg2014,Kim2014,Beugeling2014,Khodja2015,DAlessio2016,Garrison2018,Yoshizawa2018}  has emerged as a generic mechanism for thermalization in such systems. ETH has been extensively tested in numerical simulations of small quantum systems, and all known examples of  thermalizing Hermitian systems comply with ETH. In this paper, we do not distinguish between ``Thermalizable" and ``satisfying  ETH" for Hermitian systems.

 The ETH proposes that in chaotic Hermitian quantum systems, the matrix elements of simple observables display a smooth variation along the diagonal in the energy eigensates ${|E_i\rangle}$, accompanied by  fluctuations suppressed by the entropy-based exponential factor. This can be mathematically expressed as:
\begin{equation}
\langle E_i | \hat{O} | E_j \rangle = f_{\hat{O}}(E) \delta_{ij}+ g_{\hat{O}}(E,\Delta E) e^{-S(E)/2}  .
\end{equation}
In this expression, $f_{\hat{O}}$ denote smooth $O(1)$ functions, $E = \frac{E_i + E_j}{2}$, $\Delta E = |E_i - E_j|$, and    $g_{\hat{O}}$ is a function of order unity for $\Delta E=0$
that goes to zero as $ \Delta E$ becomes large. The ETH is crucial in describing thermalization, as it allows for the demonstration that starting from any initial state $\left|\psi(t)\right\rangle$ in a large system, where the coefficients $c_{\alpha}\equiv \left \langle \alpha|\psi(t)\right \rangle$ are strongly clustered around an energy $E$ (with $\left|\alpha\right\rangle$ representing the eigenstate with energy $\alpha$), the system will asymptotically approach an equilibrium state described by the microcanonical ensemble. Mathematically, this is expressed as:
\begin{equation}
\lim_{t\rightarrow \infty} \left \langle \psi(t)\right \vert \hat{O}\left \vert \psi(t)\right \rangle=\langle \hat{O}\rangle_{\text{micro}}.
\end{equation}

For the investigated  thermal quasi-Hermitian system, a similarity transformation $\mathcal{\hat{S}}_{tot}$ exists that maps it to a Hermitian system. In light of this, we introduce the following definition:

 \textit{Definition 2:} Thermalizable quasi-Hermitian system is a thermal quasi-Hermitian system without quasi-Hermiticity breaking, which  can be transformed into a Hermitian system that thermalizable. 

The equilibrium state of the quasi-Hermitian system is defined as the steady state of the thermalizable quasi-Hermitian system. In the vast majority of cases,  thermal quasi-Hermitian systems without quasi-Hermiticity breaking satisfy  definition 2, due to the fact that the transformed Hermitian system inherently includes a thermal bath. 

Next, we deduce  the equilibrium state of the thermalizable quasi-Hermitian system. The information about the equilibrium state of the quasi-Hermitian system is encoded in the expectation value of the operator $\hat{O}$. It is assumed that this operator is local enough to depend only on the degrees of freedom of the quasi-Hermitian system, i.e., $\hat{O}=\hat{O}_{S}
 \otimes\hat{I}_B$. Based on our basic assumptions in the Section II, the expectation value of $\hat{O}$ at equilibrium state  is defined as
\begin{equation}
\langle \hat{O}\rangle_{eq} \equiv  {\lim_{t\rightarrow \infty}}     \    \left \langle \psi(t)\right \vert _{\mathrm{NH}}    \hat{O}\left \vert \psi(t)\right \rangle _{\mathrm{NH}}.   
\end{equation}
Here $ \left \vert \psi(t)\right \rangle _{\mathrm{NH}} $ is the state of the thermalizable quasi-Hermitian system, and $     \left \langle \psi(t)\right \vert _{\mathrm{NH}}   \equiv (\left \vert \psi(t)\right \rangle _{\mathrm{NH}})^\dag  $.   We require the normalization of the initial state  $\left \vert \psi(0)\right \rangle _{\mathrm{NH}}  $,  and assume that the wave function coefficients $c_{\alpha}$ of the   system (determined by $   \vert \psi(0)  \rangle _{\mathrm{NH}} = c_{\alpha}\vert \alpha  \rangle_R $, where $\vert \alpha  \rangle_R$ is the self-normalized right-eigenstate of the   system's Hamiltonian with  energy $\alpha$)  are strongly clustered around a certain energy value $E$. The time evolution of the  state  $\left \vert \psi(t)\right \rangle _{\mathrm{NH}}  $ determined by  requiring $\rho(t) \equiv  \left \vert \psi(t)\right \rangle _{\mathrm{NH}}  \left \langle \psi(t)\right \vert _{\mathrm{NH}}  $ satisfies the Eq. (1).  Notice that $  \left \langle \psi(t)\right \vert_{\mathrm{NH}} \hat{O}\left \vert \psi(t)\right \rangle _{\mathrm{NH}}$ can be expressed as 
\begin{equation}
  \left \langle \psi(t)\right \vert_{\mathrm{NH}} \hat{O}\left \vert \psi(t)\right \rangle _{\mathrm{NH}} =    \frac{ \langle \widetilde{\psi(t)}  \vert_{\mathrm{NH}}   \hat{O} \vert \widetilde{\psi(t)}  \rangle _{\mathrm{NH}} }{\langle \widetilde{\psi(t)}|_{\mathrm{NH}}   | \widetilde{\psi(t)}\rangle_{\mathrm{NH}}},
\end{equation}
where $ \vert \widetilde{\psi(t)}   \rangle _{\mathrm{NH}} \equiv e^{-i\hat{H}_{tot}t} \left \vert \psi(0)\right \rangle _{\mathrm{NH}}  $, and   $ \langle \widetilde{\psi(t)} \vert _{\mathrm{NH}}   \equiv (  \vert  \widetilde{\psi(t)}  \rangle _{\mathrm{NH}})^\dag$.  In this representation, the normalization of the initial state is not required. As such,  the state of the system can be expressed as
\begin{equation}
 \vert  \widetilde{\psi(t)}  \rangle _{\mathrm{NH}} = \hat{\mathcal{S}} \otimes \hat{I}_B \left\vert \psi(t)\right\rangle_{\mathrm{0}},
\end{equation}
where $\hat{\mathcal{S}}$ represents the similarity transformation capable of constructing an operator $\mathcal{T}_c=\hat{\mathcal{S}} \hat{\mathcal{S}}^\dag$ that satisfies Lemma 1.  $\left \vert \psi(t)\right \rangle _{\mathrm{0}}\equiv e^{-i\hat{H}_{0,tot} t} \left \vert \psi(0)\right \rangle _{0}$ is the normalized states for systems with thermalizable Hermitian Hamiltonians $\hat{H}_{0,tot}= (\hat{\mathcal{S}}^{-1}  \otimes \hat{I}_B) \hat{H}_{tot} (\hat{\mathcal{S}} \otimes \hat{I}_B).$ Since similarity transformations do not alter the energy of states, under the constraint that the eigenstate thermalization hypothesis holds for the Hermitian system after the similarity transformation, the change in the magnitude of the state  with energy $\alpha$ is a smooth function $ f_{\hat{\mathcal{S}}}(\alpha)$ and does not significantly affect the relative proportions of the state's magnitude. Therefore, the wave function of the Hermitian system is also strongly clustered around the energy $E$.

Consequently, the following expression is obtained:
\begin{align}
& \lim_{t\rightarrow \infty} \langle \widetilde{\psi(t)}  \vert_{\mathrm{NH}}   \hat{O} \vert \widetilde{\psi(t)}  \rangle _{\mathrm{NH}}= \lim_{t\rightarrow \infty} \left \langle  \psi(t) \right \vert_0 \mathcal{\hat{S}}^{\dagger}\hat{O}_{S}
\mathcal{\hat{S}}\otimes \hat{I}_B\left \vert  \psi(t) \right \rangle _{{0}},   \notag    
\\
& \lim_{t\rightarrow \infty} \langle \widetilde{\psi(t)}|_{\mathrm{NH}}  |\widetilde{\psi(t)}\rangle_{\mathrm{NH}}= \lim_{t\rightarrow \infty} \left \langle  \psi(t) \right \vert_0 \mathcal{\hat{S}}^{\dagger} 
\mathcal{\hat{S}}\otimes \hat{I}_B\left \vert  \psi(t) \right \rangle _{{0}}.
\end{align}
The expression $\  \left \langle \psi(t) \right \vert_0 \mathcal{\hat{S}}^{\dagger}\hat{O}_{S}
\mathcal{\hat{S}}\otimes\hat{I}_B\left \vert \psi(t) \right \rangle _{{0}}$ is  the expectation value of the operator $\hat{O}^{\prime \ }=\mathcal{\hat{S}
}^{\dagger}\hat{O}_{S}\mathcal{\hat{S}}$ in the thermalizable Hermitian system  with Hamiltonians $\hat{H}_{0,tot}$. Similarly, 
$\left \langle \psi(t) \right \vert_0 \mathcal{\hat{S}}^{\dagger}\mathcal{\hat{S}}\otimes
\hat{I}_B\left \vert \psi(t) \right \rangle _{{0}}$ is  the expectation value of the
operator $\hat{O}^{\prime \prime }=\mathcal{\hat{S}}^{\dagger}\mathcal{\hat{S}}$ in the same
Hermitian system.  Therefore, based on our previous discussion, both limits exist.  We have
\begin{align}
\langle \hat{O}\rangle_{eq}&  =   \lim_{t\rightarrow \infty} \frac{ \langle \widetilde{\psi(t)}  \vert _{\mathrm{NH}} \hat{O} \vert \widetilde{\psi(t)}  \rangle _{\mathrm{NH}} }{\langle \widetilde{\psi(t)}| _{\mathrm{NH}}|\widetilde{\psi(t)}\rangle_{\mathrm{NH}}} \notag  \\
& = \frac{\lim_{t\rightarrow \infty} \langle \widetilde{\psi(t)}  \vert _{\mathrm{NH}} \hat{O} \vert \widetilde{\psi(t)}  \rangle _{\mathrm{NH}} }{\lim_{t\rightarrow \infty}\langle \widetilde{\psi(t)}| _{\mathrm{NH}}|\widetilde{\psi(t)}\rangle_{\mathrm{NH}}}\notag  \\
& =\frac{\langle \hat{O^{\prime}}\otimes
\hat{I}_B\rangle_{micro}}{\langle
\hat{O}^{\prime \prime }\otimes \hat{I}_B\rangle_{micro}} .
\end{align}
For any subsystem which is weakly coupled
with other parts of the system, and $\hat{O}$ is local enough to only depend
of degrees of freedom of the subsystem, i.e., $\hat{O}=\hat{O}_{sub}\otimes
\hat{I}$, we have $\langle \hat{O}\rangle_{micro}=\langle \hat{O}_{sub}
\rangle_{canonical}$ \cite{JoshuM}. Specifically, $\langle \hat{O}\rangle_{micro} =\frac{1}{\mathrm{tr}_{sub}\rho}\mathrm{tr}
_{sub}[\hat{O}_{sub}\rho]$,   where the density matrix $\rho_{sub}=e^{-\beta_{T}\hat{H}{sub}}$, with $\beta_{T}=1/T$ being the inverse temperature. We have set the Boltzmann constant to be unity.  Therefore
\begin{align}
\langle \hat{O}\rangle_{eq}
& =\frac{\langle \hat{O^{\prime}}\rangle_{canonical}}{\langle
\hat{O}^{\prime \prime }\rangle_{canonical}} \notag \\
&  =\frac{\mathrm{tr}_{s}(\hat{O^{\prime}}e^{-(\hat{\mathcal{S}}^{-1}   \hat{H}_{\mathrm{NH}} \hat{\mathcal{S}})/T})/\mathrm{tr}
_{s}(e^{- (\hat{\mathcal{S}}^{-1}   \hat{H}_{\mathrm{NH}} \hat{\mathcal{S}}) /T})}{\mathrm{tr}_{s}(\hat{O}^{\prime \prime }e^{- (\hat{\mathcal{S}}^{-1}   \hat{H}_{\mathrm{NH}} \hat{\mathcal{S}}) /T})/\mathrm{tr}
_{s}(e^{- (\hat{\mathcal{S}}^{-1}   \hat{H}_{\mathrm{NH}} \hat{\mathcal{S}}) /T})} \notag \\
&  =\frac{\mathrm{tr}_{s}(\mathcal{\hat{S}}^{\dagger}\hat{O}_{S}
\mathcal{\hat{S}}e^{-(\hat{\mathcal{S}}^{-1}   \hat{H}_{\mathrm{NH}} \hat{\mathcal{S}})/T})}{\mathrm{tr}_{s}(\mathcal{\hat{S}}^{\dagger
}\mathcal{\hat{S}}e^{-(\hat{\mathcal{S}}^{-1}   \hat{H}_{\mathrm{NH}} \hat{\mathcal{S}})/T})}\notag \\
&  =\frac{1}{\mathrm{tr}_{s}(\rho_{\mathrm{NH}})}\mathrm{tr}_{s}(\hat{O}
_{S}\rho_{\mathrm{NH}}).
\end{align}
where $\rho_{\mathrm{NH}} =e^{-\beta_{T}\hat{H}_{\mathrm{NH}}}\mathcal{T}_c$. If the Hilbert space is
limited to subsystems, the subscript  can be omitted. Therefore,
$\rho_{\mathrm{NH}}$ can be regard as the density matrix of the  quasi-Hermitian system.
This density matrix is Hermitian, i.e., $\rho_{\mathrm{NH}}=\left(
\rho_{\mathrm{NH}}\right)  ^{\dagger}.$ Then it is obvious that $\left \langle
\hat{O}\right \rangle $ is real, i.e., $\left \langle \hat{O}\right \rangle
=\left \langle \hat{O}\right \rangle ^{\ast}.$

We note that in this case, it can be demonstrated that the density matrix of any other weakly coupled Hermitian subsystem within the larger system is given by $\rho_{sub}=e^{-\beta_{T}\hat {H}_{sub}}$. This result indicates that the Hermitian component of the thermal non-Hermitian system possesses a temperature $T$. Therefore, according to definition 1, $e^{-\beta_{T}\hat{H}_{\mathrm{NH}} }\mathcal{T}_c$ can be referred to as the thermal state of a non-Hermitian system at temperature $T$. Consequently, we obtain:

\textit{Theorem 2: }The equilibrium state of thermalizable  quasi-Hermitian systems (with conserved quantity $P_c(t)\mathcal{T}_c$) at temperature $T$ is \begin{equation}\rho_{\mathrm{NH}}=e^{-\beta_{T}\hat{H}_{\mathrm{NH}}%
}\mathcal{T}_c .\end{equation}

In the derivation of Theorem 2, we have assumed that the wave function coefficients $c_{\alpha}$ are strongly clustered around a single energy value $E$. This condition seems to impose a stringent constraint on the initial state $\rho_{\mathrm{S}}(t=0)$ of the system $S$. However, in general, the system is ``much smaller than" the thermal bath, i.e., the energy scale of the system is significantly smaller than the energy scale of the thermal bath. Consequently, the initial state of the system has a insignificant impact on the overall energy distribution.  Therefore, as long as the initial state of the thermal bath is suitable (e.g., the thermal bath is initially in equilibrium at temperature $T$), the aforementioned condition can be satisfied regardless of the initial state of the system $S$. This implies that the constraints posed by the condition are not as restrictive as they may initially appear, and the results of Theorem 2 can be widely applicable.

In general, systems satisfying the  ETH exhibit ergodic behavior. It is important to note that for such systems, the time average of physical quantities over sufficiently long periods must be unique. This is because the evolution from different initial states can always be considered as the evolution of the same state with varying initial times.  Consequently, there must be only one $\mathcal{T}_c$ (up to a multiplicative factor. A more rigorous discussion regarding this assertion is provided in Appendix C), which  is uniquely determined by the coupling $\hat{C}_{a}$.

\subsection{Restriction of relaxation time on equilibrium state} 
In discussing the equilibrium states of quasi-Hermitian systems with finite relaxation times, it is important  to investigate their characteristic properties. The states presented above correspond to \emph{non-Boltzmann distributions}. It is well-established that for Hermitian systems, the weights of distinct quantum states adhere to the Boltzmann distribution, expressed as $\rho = e^{-\beta_{T}\hat{H}}$. However, for non-Hermitian   systems, the conventional Boltzmann distribution transforms into a non-Boltzmann distribution, denoted by $\rho_{\mathrm{NH}} = e^{-\beta_{T}\hat{H}_{\mathrm{NH}}}\mathcal{T}_c$. It is essential to note that the Boltzmann distribution, constructed under the assumption that the probability of states with energy $E_{n}$ is $P_{n} \propto e^{-\beta_{T}E_{n}}$, is generally not satisfied. The probability of states with energy $E_{n}$ is $P^\mathrm{NH}_{n} \propto e^{-\beta_{T}E_{n}}W_n \neq e^{-\beta_{T}E_{n}}$, where $W_n = \langle E_n |_{L} \mathcal{T}_c | E_n \rangle_{L}$, with $ | E_n \rangle_{L}$ is the left-eigenstate   which biorthogonal to self-normalized $\left \vert E_n \right \rangle _{R}$. To characterize the deviation from the Boltzmann distribution, we define the abnormal index as:
\begin{equation}
 A = \ln \langle \mathcal{T}_c \rangle_{\rho_{\mathrm{NH}}},
\end{equation}
which $  \sim \ln (\sum_n W_n P^\mathrm{NH}_n)$, where $\langle   \   \    \rangle_{\rho_{\mathrm{NH}}}$ denotes the expectation value concerning the density matrix $\rho_{\mathrm{NH}}$. We select the average value of $W_n$ to be $1$, or $\mathrm{tr} \  \mathcal{T}_c = \mathrm{dim} \  \mathcal{T}_c$. If $A < 0$, it suggests that the system prefers occupying higher energy states than the Boltzmann distribution predicts. Conversely, if $A > 0$, the system tends to favor lower energy states.

By necessity, the distribution of non-Hermitian systems should reduce to the Boltzmann distribution under the Hermitian limit. However, it appears that the distribution in Theorem 2 is not reducible to the Boltzmann distribution---according to prior result, for systems in the Hermitian limit, a set of equilibrium states $\rho  = e^{-\beta_{T}\hat{H}{{}}}\mathcal{T}_c$ seems to exist, which deviate from the standard Boltzmann distribution $\rho  = e^{-\beta_{T}\hat{H}_{{}}}$.

The key point is that the equilibrium state provided by Theorem 2 represents a physically plausible equilibrium state. Nonetheless, numerous quasi-Hermitian systems display divergent relaxation times in the equilibrium state. For instance, under the Hermitian limit, if $\mathcal{T}_c$ is unequal to the identity matrix $\hat{I}$, $P_c(t)\mathcal{T}_c$ is reduced from a path-dependent conserved quantity to an ordinary conserved quantity. Unlike path-dependent quantities, the presence of conserved quantities can impede system thermalization, resulting in the system's final state preserving the same physical quantity as the initial state, whereas the final state of a thermalizable system should be independent of the initial state.

More specifically, if finite relaxation time is required, the existence of path-dependent conserved quantities imposes a certain constraint between the non-Hermitian strength of the system and the abnormal index of the equilibrium state. Utilizing the expression of conserved quantities, we can estimate:
\begin{equation}
\left \langle \mathcal{T}_c\right \rangle_{\rho_{0}} \sim e^A e^{\gamma \tau},
\end{equation}
where $\gamma \equiv \sqrt{\mathrm{tr} \varUpsilon^2/\dim \varUpsilon }$ represents the non-Hermitian strength of the system, and $\tau$ denotes the relaxation time of the system. This leads to:
\begin{equation}
\tau \sim |\frac{A}{\gamma}|.
\end{equation}
If its relaxation time $\tau$ diverges, then such a state cannot be experimentally achieved. Based on this observation, we propose the following theorem:

\textit{Theorem 3 -- } For a thermalizable quasi-Hermitian system, if its equilibrium states exhibit finite relaxation times, its abnormal index $A$ cannot be significantly larger than the system's non-Hermitian strength $\gamma$.
 
\subsection{Summary} In this section, we briefly summarize the three theorems of the theory of quasi-Hermitian systems and their relationships. Quasi-Hermitian systems are inherently unstable; Theorem 1 provides the necessary physical conditions for achieving stability in such systems. Lemma 1 offers a more technical representation of this theorem, which is advantageous for theoretical applications. Next, Theorem 2 describes the physically allowable equilibrium states of a stable quasi-Hermitian system. It is essential to note that due to the inherent characteristics of non-Hermitian systems, these equilibrium states often possess exceedingly long relaxation times. As a result, they become experimentally unobservable; Theorem 3 provides a method to exclude these unobservable states, enabling a focus on practically observable states in quasi-Hermitian systems. In conclusion, these three theorems offer a comprehensive understanding of the stability, equilibrium states, and relaxation times of quasi-Hermitian systems.

\section{Quantum Statistical Mechanics for non-Hermitian systems}

\bigskip

For non-Hermitian Hamiltonians possessing complex eigenvalues, the eigenvalues with negative imaginary components demonstrate dissipation at a rate proportional to the imaginary part of the corresponding eigenstate (we have adjust the maximum imaginary part of the eigenvalue of the non-Hermitian system to zero). 
During the evolution of time, in the weak coupling limit, these dissipative states will decay rapidly (compared to the timescale of non-dissipative state evolution), and hence do not significantly impact the dynamics of non-dissipative states. Consequently, the steady state of the system can be well described by the Hamiltonian restricted to the subspace of non-dissipative states. A more detailed technical discussion based on the master equation is presented in Appendix D.

Therefore, in the case of a non-Hermitian system, the original  model is reduced to a subspace consisting of m-fold degenerate quantum states that share the same maximum imaginary part among all eigenvalues. Specifically, the Hamiltonian of the system and the coupling between the system and the bath restricted to this subspace  are:
\begin{equation}
\hat{H}^R_{\mathrm{NH}}=P\hat{H}_{\mathrm{NH}}P,
\end{equation}
\begin{equation}
\hat{H}_{BS}^{R}=\sum_{a}\lambda_a[( P \hat{C}_{a} P )\otimes \hat{B}_{a}+( P \hat{C}^\dag_{a} P)\otimes \hat{B}^\dag_{a}] .
\end{equation}
Where $P=\sum_{Im E_{m} =0}\  \left \vert m\right \rangle _{R}\left \langle
m\right \vert _{L}$ is the projection operator to the subspace with zero
imaginary part. We call this system as \textit{reduced thermal quasi-Hermitian
system.} We defined a thermal non-Hermitian system is thermalizable if its corresponding reduced thermal quasi-Hermitian system is thermalizable. Therefore,  studying the equilibrium state of a non-Hermitian system only requires investigating the equilibrium state of the reduced thermal quasi-Hermitian system.

\subsection{ Equilibrium state of non-Hermitian systems}

The only difference between the  reduced thermal quasi-Hermitian system and the thermal quasi-Hermitian system introduced in the previous section is that $ P\hat{C}_{a} P $ and $ P\hat{C}^\dag_{a}P$ are not necessarily Hermitian conjugates. However, after repeated deduction, we find that Theorems 1, 2 and 3 still hold. Therefore, we also can employ Theorems 1, 2, and 3 to investigate the reduced quasi-Hermitian system and obtain the equilibrium state of the non-Hermitian system.   Only the form of Lemma 1 has changed a little. Condition (1) in Lemma 1 is changed to $P\hat{C}_{a}P\mathcal{T}_c^{{R}}\mathcal{-T}_c^{{R}} P^{\dag} \hat{C}_{a}P^{\dag}
 =0$ for all $a$. Where $\mathcal{T}_c^{{R}} $ is a positive definite Hermitian operator defined in the Hilbert space where the reduced quasi-Hermitian system is located.

 We note that the reduction of the equilibrium state of a non-Hermitian system to the equilibrium state of a reduced quasi-Hermitian system is only valid in the regime of sufficiently weak coupling. This approximation relies on the dissipative time scale $\tau_{diss}$ being much smaller than the time scale of evolution for non-dissipative states $\tau_{nondiss}$. In fact, it will be shown in Appendix D that the error $\delta$ of this algorithm is approximately 
$\delta \sim\tau_{diss}/\tau_{nondiss}. $
Therefore, this equilibrium state is actually an approximation. Only when the conserved quantity in Theorem 1 exists, non-Hermitian systems have an exact equilibrium state.

\subsection{Role of conserved quantity in equilibrium state of non-Hermitian systems}

In the context of a physical perspective, the emergence of this approximate equilibrium state is intimately connected to the existence of conserved quantities. While the system initially lacks exact conserved quantities, the dissipative state's dissipation leads to the rapid appearance of an approximate conserved quantity $P_c^R(t)\mathcal{T}_c^{R}$ on the time scale $\tau_{diss}$. Notably, the system's thermalization process has barely commenced on this time scale. This implies that the thermalization process is consistently protected by the conserved quantity, resulting in a stability of non-Hermitian system. The conserved quantity in conjunction with the effective transitions within the subspace $\hat{H}^R_{\mathrm{NH}}$ co-determines the system's equilibrium state.

In addition, we point out that projecting the system and ascertaining whether the reduced thermal quasi-Hermitian system satisfies Theorem 1 indeed furnishes a method for assessing the presence of the emerged conserved quantity $P_c^R(t)\mathcal{T}_c^{R}$ within the original system.

\subsection{Thermodynamic quantities of non-Hermitian systems} 
 
In the field of statistical mechanics for Hermitian systems, partition function $Z=\mathrm{tr}\ \rho$, and its consequential free energy $F = -T\ln Z$ serve as insightful tools for effectively calculating physical quantities. Similarly, in the realm of non-Hermitian systems, these elements may play equally substantial roles.

Hence, we propose the partition function for non-Hermitian systems as:
\begin{equation}
Z(\hat{H}_{\mathrm{NH}}, T, \mathcal{T}_c) = \mathrm{tr}(e^{-\frac{1}{T}\hat{H}^R_{\mathrm{NH}}} \mathcal{T}_c^R).
\end{equation}
With the corresponding free energy given by:
\begin{equation}
F(\hat{H}_{\mathrm{NH}}, T, \mathcal{T}_c) = -T \ln Z(\hat{H}_{\mathrm{NH}}, T, \mathcal{T}_c).
\end{equation}
Having defined these components, we can proceed to discuss the definitions of thermodynamic quantities in non-Hermitian systems, and their associations with partition functions and free energy.

The thermodynamic average of physical quantities, which predominantly appear as derivatives of the Hamiltonian concerning specific parameters, is initially discussed. In non-Hermitian systems, for a generic parameterised operator $\hat{O} = \frac{\partial \hat{H}_{\mathrm{NH}}(\lambda)}{\partial \lambda}$, it is established that:
\begin{equation}
\langle \hat{O} \rangle \equiv \frac{1}{Z} \mathrm{tr}(\hat{O} \rho_{\mathrm{NH}}) = -T \frac{\partial \ln Z(\hat{H}_{\mathrm{NH}}(\lambda), T, \mathcal{T}_c)}{\partial \lambda},
\end{equation}
And equilevanty that,
\begin{equation}
\langle \hat{O}\rangle = \frac{\partial F(\hat{H}_{\mathrm{NH}}(\lambda), T, \mathcal{T}_c)}{\partial \lambda}.
\end{equation}
These equations provide a well-rounded framework for evaluating both Hermitian and non-Hermitian systems.

Next, we consider energy. The   internal energy is generated by weighting each energy state by its probability, culminating in:
\begin{equation}
E = \sum_n E_n P_n = \frac{1}{Z} \mathrm{tr}(\hat{H}^R_{\mathrm{NH}} \rho_{\mathrm{NH}}) = F - T \frac{\partial F}{\partial T}.
\end{equation}
In this context, energy is defined as the real part of an eigenstate's eigenvalue, with the imaginary part representing the state's dissipation rate. Similar to the Hermitian case, energy can be formulated via the Legendre transformation of free energy.

Lastly, entropy in non-Hermitian systems is presented as the von Neumann entropy:
\begin{equation}
S = -\mathrm{tr}(\bar{\rho}_{\mathrm{NH}} \ln \bar{\rho}_{\mathrm{NH}}),
\end{equation}
where $\bar{\rho}_{\mathrm{NH}}=\frac{\rho_{\mathrm{NH}}}{\mathrm{tr}\rho_{\mathrm{NH}}}$. The entropy measures the degree of disorder or uncertainty inherent in a system. Notably, in non-Hermitian systems, entropy is not equivalent to the derivative of free energy concerning temperature, i.e., $S \neq -\frac{\partial F}{\partial T}$. This deviation highlights the distinctive attributes of non-Hermitian systems.

\section{Example}
\subsection{A two-level system}
We consider a sample two-level non-Hermitian system coupled to a thermal bath, featuring a spinless fermion distributed across two lattice sites. The Hamiltonian of this non-Hermitian system   is given by
\begin{equation}
\hat{H}_{\mathrm{NH}}=\sigma_{x}+i\gamma \sigma_{y}%
\end{equation}
on the pseudo-spin space $(%
\begin{array}
[c]{c}%
\left \vert \uparrow \right \rangle \\
\left \vert \downarrow \right \rangle
\end{array}
)$, where $\left \vert \uparrow \right \rangle$ and $\left \vert \downarrow \right \rangle$ represent the quantum states of a single fermion on site $1$ and site $2$, respectively.
The coupling between system $\mathrm{S}$ and bath $\mathrm{B}$ is captured by the term $\hat{H}_{BS}=\sum_{a}\gamma_{a}\hat{C}_{a}\otimes \hat{B}_{a}$. Here, $\hat{B}_{a}$ and $\hat{C}_{a}$ denote the operators of systems $\mathrm{B}$ and $\mathrm{S}$, respectively. We define $\hat{C}_{a}=\hat{n}_{a}$ as the particle number operator of $S$ on lattice site $a$, with $a=1,2$. We assume that $\gamma_{a} \ll \gamma,1$ for our analysis.   This system can be implemented by a controlled open quantum system \textrm{S} coupling to two
environments \textrm{B} and \textrm{E} \cite{QWG3}.

 For the non-Hermitian Hamiltonian $\hat{H}_{\mathrm{NH}}$, at $\gamma = 1$, a typical $\mathcal{PT}$-symmetry spontaneous breaking  occurs: For the case $  \gamma<1$, the energy levels for states $\left \vert +\right \rangle_R$ and $\left \vert -\right \rangle_R$ are $E_{\pm}=\pm \sqrt{1-\gamma^2}$. For the case $ \gamma>1$, the energy levels for states $\left \vert +\right \rangle_R$ and $\left \vert -\right \rangle_R$ are $E_{\pm}=\pm i\sqrt{\gamma^{2}-1}$.  For the case $\gamma=1$, the system is at the exceptional point  with state coalescing and energy degeneracy, i.e., $E_{\pm}=0$.

Firstly, we consider the $\mathcal{PT}$-symmetric breaking phase. Due to the system  has a complex energy spectrum, we need to project the system into the subspace where the eigenstate with the largest imaginary part is located. The projection operator can be expressed as: \begin{equation} P= \left \vert +\right \rangle_R \left \langle + \right \vert_L, \end{equation} 
where $\left \vert +\right \rangle_L$ is the eigenstate of Hamiltonian $\hat{H}^\dag_{\mathrm{NH}}$  which satisfies $_L\left \langle + | +\right \rangle_R=1$. Acting this projection operator  on the Hamiltonian and the coupling terms, we get the following result: 
\begin{equation} \hat{H}^R_{\mathrm{NH}}= 0, \end{equation}
 \begin{equation} \hat{H}^R_{BS}=\sum_{a}\gamma_{a} \left \langle + \right \vert_L \hat{n}_{a} \left \vert +\right \rangle_R ( \left \vert +\right \rangle_R \left \langle + \right \vert_L ) \otimes \hat{B}_{a}. \end{equation} Here, $\hat{H}^R_{\mathrm{NH}}$ and $\hat{H}^R_{BS}$ represent the projected Hamiltonian of the system and the projected coupling term between system $\mathrm{S}$ and bath $\mathrm{B}$, respectively.
According to the expression of $\hat{H}^R_{\mathrm{NH}}$  and $\hat{H}^R_{BS}$, we can determine the reduced conserved quantity $\mathcal{T}_c^{R}$ is
\begin{equation}
\mathcal{T}_c^{R}= \left \vert +\right \rangle _R  \left  \langle + \right \vert_R =\frac{1}{2}(I+ \sqrt{1-\gamma^{-2}}\sigma_y + \gamma^{-1} \sigma_z).
\end{equation}
By using 
$\rho_{\mathrm{NH}}=e^{-\beta_{T}%
 \hat{H}^R_{\mathrm{NH}} }\mathcal{T}_c^{R}$, we have
\begin{equation}
\rho_{\mathrm{NH}}=  \frac{1}{2}(I+ \sqrt{1-\gamma^{-2}}\sigma_y + \gamma^{-1} \sigma_z).
\end{equation}
The expectation value of $ \left \langle \vec{\sigma} \right \rangle \equiv (\left \langle  \sigma_x  \right \rangle, \left \langle \sigma_y  \right \rangle, \left \langle \sigma_z  \right \rangle)$ is calculate as
\begin{equation}
 \left \langle \vec{\sigma} \right \rangle =(0,\sqrt{1-\gamma^{-2}},\gamma^{-1}).
\end{equation}
This result can also be obtained from the partition function.
The  partition function of the system is
\begin{equation}
Z(\vec{\lambda})= \mathrm{tr} (e^{-\frac{1}{T}(\hat{H}^R_{\mathrm{NH}}+\vec{\lambda} \cdot  \vec{\sigma}) 
 }\mathcal{T}_c^R),
\end{equation}
where $\vec{\lambda} \cdot  \vec{\sigma}= \lambda_x \sigma_x+\lambda_y \sigma_y+\lambda_z \sigma_z $.
The expectation value is 
 \begin{equation}
 \left \langle \vec{\sigma} \right \rangle=-T \nabla_{\vec{\lambda}} \ln Z(\vec{\lambda} )|_{\vec{\lambda}=0}=(0,\sqrt{1-\gamma^{-2}},\gamma^{-1}).
\end{equation}

Next, we consider the  $\mathcal{PT}$-symmetry phase.  For this system, the $\mathcal{T}_c$ is
\begin{equation}
\mathcal{T}_c=\ I+ \gamma \sigma_{z}.
\end{equation}
In the pseudo-spin space, this is equivalent to a rotational symmetry along the $z$ direction. In the original representation, the symmetry necessitates that the operator can be expressed as a sum of local operators. Specifically, since any local operator adheres to this symmetry, it is quite natural for the coupling to conform to it as well. The equilibrium state
protect by $\mathcal{T}_c$ is%
\begin{equation}
\rho_{\mathrm{NH}}=\frac{1}{2}\ I\ +\frac{\gamma}{2}\sigma_{z}-\frac{1}%
{2}\sqrt{1-\gamma^{2}}\frac{e^{2\sqrt{1-\gamma^2}/T}-1}{e^{2\sqrt{1-\gamma^2}/T}+1}%
\sigma_{x}.
\end{equation}
The expected value $\left \langle \vec{\sigma} \right \rangle$  is calculate as
\begin{equation}
\left \langle \vec{\sigma} \right \rangle=(-\sqrt{1-\gamma^{2}}\frac{e^{2\sqrt{1-\gamma^2}/T}-1}{e^{2\sqrt{1-\gamma^2}/T}+1},0,\gamma).
\end{equation}
Similar to the $\mathcal{PT}$-symmetric phase, this result can also be given by the partition function.
The  partition function of the system is
\begin{equation}
Z(\vec{\lambda})= \mathrm{tr} (e^{-\frac{1}{T}(\hat{H}_{\mathrm{NH}}+\vec{\lambda} \cdot  \vec{\sigma}) 
 }\mathcal{T}_c),
\end{equation}
and the expected value $\left \langle \vec{\sigma} \right \rangle$  is calculate as
 \begin{equation}
  \left \langle \vec{\sigma} \right \rangle=-T \nabla_{\vec{\lambda}} \ln Z(\vec{\lambda} )|_{\vec{\lambda}=0}=(-\sqrt{1-\gamma^{2}}\frac{e^{2\sqrt{1-\gamma^2}/T}-1}{e^{2\sqrt{1-\gamma^2}/T}+1},0,\gamma).
\end{equation}
The system exhibits abnormal index with $A=\ln(1+\gamma^2) $. To determine if these equilibrium states are restricted, we can compute $A/\gamma$. It is discovered that the value will not significantly exceed 1, which implies that these equilibrium states can be experimentally observed.

 \begin{figure}[ptb]
\includegraphics[width=9cm ]{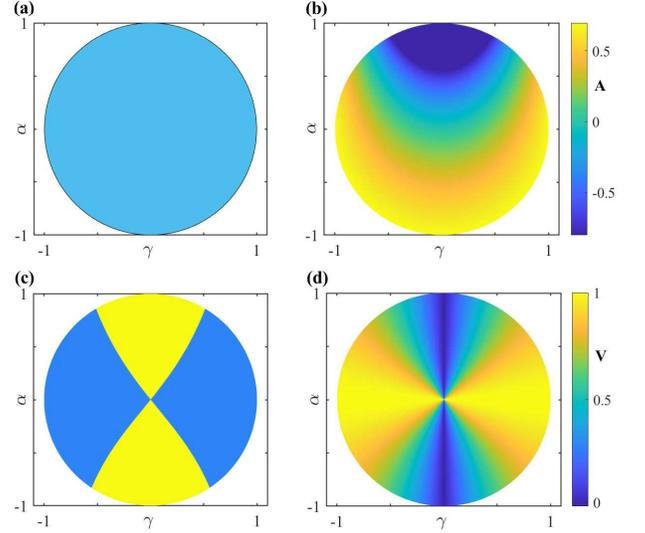}
\caption{ (a) Permissible values of $\alpha$ as a function of non-Hermitian strength $\gamma$.
(b) Equilibrium state anomalous index $A$ for varying $\alpha$ and $\gamma$ values. We set $T=0.1$.
(c) Acceptable $\alpha$ values (blue region) as a function of non-Hermitian strength $\gamma$ considering relaxation time constraints. The area with $A/\gamma > 1.25$ (yellow region) is excluded.
(d) Convergence rate $V$ to the equilibrium state for a given coupling $\hat{H}_{BS}$, with the convergence rate of the Hermitian case set to 1. }
\end{figure}

\subsection{Equilibrium state protected by other conserved quantity}
In this subsection, we  do not limit  the coupling between the system and the thermal bath, consider all possible equilibrium states of the system, and judge which equilibrium states have too long relaxation time to be observed in the experiment, as an application of Theorem 3. Since the conserved quantity in $\mathcal{PT}$-symmetric breaking phase is trivial, we will only consider the $\mathcal{PT}$-symmetric phase in this subsection.

Through direct calculation, we obtain  $\mathcal{T}_c$ satisfying  $\hat{H}_{\mathrm{NH}}\mathcal{T}_c-\mathcal{T}_c\hat{H}_{\mathrm{NH}}^{\dagger}=0$
 can be expressed as  
\begin{equation}
\mathcal{T}_c=I+\gamma\sigma_z+ \alpha \sigma_x,
\end{equation}
where $ -\sqrt{1-\gamma^2} \leq \alpha  \leq  \sqrt{1-\gamma^2}$. As shown in Fig. 2(a).
  We can
get the corresponding equilibrium state
\begin{equation}
\rho_{\mathrm{NH}}=\frac{1}{2}\ I\ +\frac{\gamma}{2}\sigma_{z}-\frac{1}%
{2}\sqrt{1-\gamma^{2}}\frac{  ke^{2\sqrt{1-\gamma^2}/T}-1}{ke^{2\sqrt{1-\gamma^2}/T }+1}\sigma_{x}.
\end{equation}
where $k=    \frac{ \sqrt{1-\gamma^2}-\alpha}{\sqrt{1-\gamma^2}+\alpha} 
$.
The expectation value  $\left \langle \vec{\sigma} \right \rangle$  is calculate as
\begin{equation}
\left \langle \vec{\sigma} \right \rangle=(-\sqrt{1-\gamma^{2}}\frac{ke^{2\sqrt{1-\gamma^2}/T}-1}{ke^{2\sqrt{1-\gamma^2}/T}+1},0,\gamma),
\end{equation}
as seen in Fig. 3. The same result can also be obtained by using partition function method.
 This distribution has an abnormal index  
\begin{equation}
A= \ln [ 1+ \gamma^2 - \alpha  \sqrt{1-\gamma^{2}} \frac{ke^{2\sqrt{1-\gamma^2}/T}-1}{ke^{2\sqrt{1-\gamma^2}/T}+1}],
\end{equation}
as seen in Fig. 2(b). We calculated the values of $A/\gamma$  and found that they are larger near the y-axis, suggesting that the equilibrium  states corresponding to these conserved quantities should be excluded from consideration, see in Fig. 2(c).

\begin{figure}[ptb]
 
\includegraphics[width=9cm]{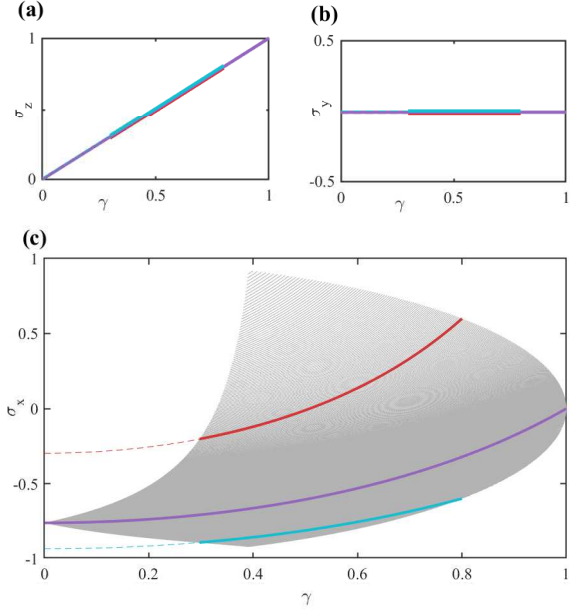}
\caption{(a) Expectation value of $\sigma_z$ as a function of $\gamma$ for $T=1$.
(b) Expectation value of $\sigma_y$ as a function of $\gamma$ for $T=1$.
(c) Expectation value of $\sigma_x$ as a function of $\gamma$ with a shaded region indicating the relaxation time is within acceptable limits (relaxation time not exceeding twice the Hermitian case) for $T=1$.
In the figure, red, purple, and cyan lines correspond to $\alpha=-0.6, 0,0.6$, respectively. The dashed line signifies the system condition where the relaxation time exceeds twice the Hermitian case.}
\end{figure}
To verify this result, we construct a coupling that satisfies the symmetry $\mathcal{T}_c(\gamma,\alpha)$ and calculate the relaxation time. For a system with
parameter ($\gamma$,$\alpha$), a possible coupling is 
\begin{equation}
\hat{H}_{BS} = \frac{\gamma_{1}}{2}\left(
\begin{array}
[c]{cc}%
1 + \frac{1}{ \sqrt{\left(\frac{\alpha}{\gamma}\right)^{2}+1}} & \frac{1}{\sqrt{\left(\frac{\gamma}{\alpha}\right)^{2}+1}}\\
\frac{1}{\sqrt{\left(\frac{\gamma}{\alpha}\right)^{2}+1}} & 1 - \frac{1}{ \sqrt{\left(\frac{\alpha}{\gamma}\right)^{2}+1}}%
\end{array}
\right) \otimes \hat{B}_{1}.
\end{equation}

We   compute the time evolution and obtain the relaxation time of such a system, which is proportional to $\sqrt{(\frac{\alpha}{\gamma\sqrt{1+\alpha^2}})^2+1} \sim A/\gamma$. As shown in Fig. 2(d), the region exhibiting longer relaxation times, as determined by this direct calculation, aligns well with the area derived from Theorem 3.

\section{Conclusions}
 In conclusion, our systematic investigation of statistical mechanics for non-Hermitian quantum systems has revealed valuable insights into their thermodynamic properties and stability conditions. Our analysis has demonstrated that the stability of these systems requires the existence of a single path-dependent conserved quantity, be it exact or approximate, which, when combined with the system's Hamiltonian, governs the equilibrium state. By identifying the relationship between the Hamiltonian and the associated conserved quantity, we have proposed criteria for discerning equilibrium states with finite relaxation times.

Our investigation provides important insights into understanding non-Hermitian systems in practical contexts, establishing the foundation for potential applications of quantum non-Hermitian  systems. Future research can build on our findings by exploring more complex systems. Adding to this, our exploration of the statistical mechanics associated with non-Hermitian systems sets a firm groundwork for forthcoming research into their thermodynamic behaviour. A distinctive aspect is that non-Hermitian systems, due to their non-Boltzmann distribution, are expected to exhibit unique thermodynamic properties. Therefore, a deepened understanding of these properties could facilitate more nuanced applications. Additionally, our research points towards an intriguing possibility - that changes in the conserved quantities during tuning of the system's Hamiltonian could give rise to new singularities within the partition function, potentially ushering in unprecedented thermodynamic phase transitions. This prospect presents an exhilarating avenue for future exploration, broadening the scope of research within this fascinating domain.

\acknowledgments This work was supported by the Natural Science Foundation of China (Grants No. 11974053 and No. 12174030). We are grateful to  Shi-Qi Zhao for helpful discussions. 

\section*{ Appendix A: Master equation of quasi-Hermitian system}
\setcounter{equation}{0}
\renewcommand\theequation{A\arabic{equation}}
Generally speaking, the derivation of a quantum Markovian master equation is
performed in the interaction picture. Thus, we write the time evolution
equation of the density matrix of system and the bath as in the interaction
picture \cite{Lindblad76,GKS76,BreuerPetruccione}
\begin{equation}
\frac{d}{dt}\rho^{I}_{\mathrm{S+B}}(t)=-i(\hat{V}_{I}\left(  t\right)
\rho^{I}_{\mathrm{S+B}}(t)-\rho^{I}_{\mathrm{S+B}}(t)\hat{V}_{I}^{\dag}\left(
t\right)  ), \label{2}%
\end{equation}
where $\rho^{I}_{\mathrm{S+B}}(t)=e^{i\hat{H}_{\mathrm{eff},0}t}%
\rho_{\mathrm{S+B}}(t)e^{-i\hat{H}_{\mathrm{eff},0}^{\dag}t}$ and $\hat{V}%
_{I}\left(  t\right)  =e^{i\hat{H}_{\mathrm{eff},0}t}\hat{H}_{BS}e^{-i\hat{H}_{\mathrm{eff},0}t}$ is the interaction Hamiltonian in the
interaction picture and $\hat{H}_{\mathrm{eff},0}=\hat{H}_{\mathrm{NH}}%
\otimes \hat{I}_{B}+\hat{I}_{S}\otimes \hat{H}_{B}$. Here $\rho_{\mathrm{S+B}} $ is the (non-normalized) density matrix of thermal bath and system
Its equivalent integral form is
\begin{equation}
\rho^{I}_{\mathrm{S+B}}(t)=\rho^{I}_{\mathrm{S+B}}(0)-i\int_{0}^{t}ds\left[
\hat{V}_{I}\left(  s\right)  \rho^{I}_{\mathrm{S+B}}\left(  s\right)
-\rho^{I}_{\mathrm{S+B}}\left(  s\right)  \hat{V}_{I}^{\dag}\left(  s\right)
\right]  . \label{3}%
\end{equation}
Substitute Eq. (\ref{3}) into Eq. (\ref{2}), we obtain \begin{widetext}
\begin{align}
\frac{d}{dt}\rho^{I}_{\mathrm{S+B}}(t)  &  =-i\left[  \hat{V}_{I}\left(
t\right)  \rho^{I}_{\mathrm{S+B}}(0)-\rho^{I}_{\mathrm{S+B}}(0)\hat{V}%
_{I}^{\dag}\left(  t\right)  \right]  -  \bigg \{ \hat{V}_{I}\left(  t\right)  \int_{0}^{t}ds\left[  \hat{V}%
_{I}\left(  s\right)  \rho^{I}_{\mathrm{S+B}}\left(  s\right)  -\rho
^{I}_{\mathrm{S+B}}\left(  s\right)  \hat{V}_{I}^{\dag}\left(  s\right)
\right] \nonumber \\
&    -\int_{0}^{t}ds\left[  \hat{V}_{I}\left(  s\right)  \rho^{I}_{\mathrm{S+B}%
}\left(  s\right)  -\rho^{I}_{\mathrm{S+B}}\left(  s\right)  \hat{V}_{I}%
^{\dag}\left(  s\right)  \right]  \hat{V}_{I}^{\dag}\left(  t\right)
\bigg \}.
\end{align}
\end{widetext}

After taking the partial trace over the degrees of freedom of thermal bath
\textrm{B} and use the Born approximation $\mathrm{tr}_{\mathrm{B}}\left[  \hat{V}%
_{I}\left(  t\right)  \rho^{I}_{\mathrm{S+B}}(0)-\rho^{I}_{\mathrm{S+B}%
}(0)\hat{V}_{I}^{\dag}\left(  t\right)  \right]  =0$, $\rho^{I}_{\mathrm{S+B}%
}(s)\sim \rho^{I}_{\mathrm{S}}(s)\otimes \rho^{I}_{\mathrm{B}}$, the Markov
approximation $\rho^{I}_{\mathrm{S}}(s)\sim \rho^{I}_{\mathrm{S}}(t)$, we have
\begin{widetext}
\begin{align}
\frac{d}{dt}\rho^{I}_{\mathrm{S}}(t)  &  =-\mathrm{tr}_{\mathrm{B}}\{ \hat{V}%
_{I}\left(  t\right)  \int_{0}^{t}ds\left[  \hat{V}_{I}\left(  s\right)
\left(  \rho^{I}_{\mathrm{S}}(t)\otimes \rho^{I}_{\mathrm{B}}\right)  -\left(
\rho^{I}_{\mathrm{S}}(t)\otimes \rho^{I}_{\mathrm{B}}\right)  \hat{V}_{I}%
^{\dag}\left(  s\right)  \right] \nonumber \\
&  -\int_{0}^{t}ds\left[  \hat{V}_{I}\left(  s\right)  \left(  \rho
^{I}_{\mathrm{S}}(t)\otimes \rho^{I}_{\mathrm{B}}\right)  -\left(  \rho
^{I}_{\mathrm{S}}(t)\otimes \rho^{I}_{\mathrm{B}}\right)  \hat{V}_{I}^{\dag
}\left(  s\right)  \right]  \hat{V}_{I}^{\dag}\left(  t\right)  \}.
\end{align}
\end{widetext}
We substitute $s$ by $t-s$, then the above equation can be expressed as%
\begin{align}
\frac{d}{dt}\rho^{I}_{\mathrm{S}}(t)  &  =\mathrm{tr}_{\mathrm{B}}[\int_{0}^{t}%
ds\hat{V}_{I}\left(  t-s\right)  \rho^{I}_{\mathrm{S}}(t)\otimes \rho
^{I}_{\mathrm{B}}\hat{V}_{I}^{\dag}\left(  t\right) \nonumber \label{5}\\
&  -\int_{0}^{t}ds\hat{V}_{I}\left(  t\right)  \hat{V}_{I}\left(  t-s\right)
\rho^{I}_{\mathrm{S}}(t)\otimes \rho^{I}_{\mathrm{B}}]+H.c.
\end{align}

For the interaction Hamiltonian $\hat{V}_{I}\left(  t\right)  =e^{i\hat
{H}_{\mathrm{eff},0}t}\hat{H}_{BS}e^{-i\hat{H}_{\mathrm{eff},0}t}$,
we insert the identity operator $\hat{I}=\sum_{m}\left \vert m\right \rangle
_{R}\left \langle m\right \vert _{L}\otimes \hat{I}_{B}$ in it and get
\begin{widetext}
\begin{align}
\hat{V}_{I}\left(  t\right)   &  =e^{i\hat{H}_{\mathrm{eff},0}t}\left[
\sum_{a,b}\sum_{m}\left \vert m\right \rangle _{R}\left \langle m\right \vert
_{L}\gamma_a \hat{C}_{a}\sum_{n}\left \vert n\right \rangle
_{R}\left \langle n\right \vert _{L}\otimes \hat{B}_{a}\right]  e^{-i\hat
{H}_{\mathrm{eff},0}t}\nonumber \\
&  =\sum_{a,b}\sum_{m}\sum_{\omega}e^{-i\omega t}\left \vert m\right \rangle
_{R}\left \langle m\right \vert _{L}\gamma_a \hat{C}_{a}\left \vert
m+\omega \right \rangle _{R}\left \langle m+\omega \right \vert _{L}\otimes \hat
{B}_{a}\left(  t\right)  ,
\end{align}
\end{widetext}
where $\left \vert m\right \rangle _{R}/\left \vert m\right \rangle _{L}$ is the
right/left eigenstate of $\hat{H}_{\mathrm{NH}}$ with eigenvalue $E_{m}$.
$\omega=E_{n}-E_{m}$, and $\left \vert m+\omega \right \rangle $ means the state
with energy $E_{n}=E_{m}+\omega$. We have utilized an alternative general expression for $\hat{H}_{BS}$, where $\hat{H}_{BS} = \sum_{a} \gamma_{a} \hat{C}_{a} \otimes \hat{B}_{a}$. In this case, $\gamma_{a}$ are real
numbers, and $\hat{C}_{a}$, $\hat{B}_{a}$ are Hermitian operators. Here we
define a non-Hermitian operator%
\begin{equation}
\label{7}\hat{A}_{a}(\omega)=\sum_{m}\left \vert m\right \rangle _{R}%
\left \langle m\right \vert _{L}\gamma_{a} \hat{C}_{a}\left \vert m+\omega
\right \rangle _{R}\left \langle m+\omega \right \vert _{L},
\end{equation}
where $a=1,2,3,..., L$. Then $\hat{V}_{I}\left(  t\right)  $ can be written
as
\begin{equation}
\hat{V}_{I}\left(  t\right)  =\sum_{a}\sum_{\omega}e^{-i\omega t}\hat{A}%
_{a}(\omega)\otimes \hat{B}_{a}(t).
\end{equation}
Substituting this form of $\hat{V}_{I}(t)$ to Eq.(\ref{5}), we get
\begin{widetext}
\begin{align}
\frac{d}{dt}\rho^{I}_{\mathrm{S}}(t)  &  =\sum_{a,b}\sum_{\omega,\omega_{1}}e^{i\left(  \omega_{1}-\omega \right)  t}\hat{A}_{b}%
(\omega)\rho^{I}_{\mathrm{S}}(t)\hat{A}_{a}^{\dag}(\omega_{1})\int_{0}%
^{t}dse^{i\omega s}\mathrm{tr}_{\mathrm{B}}\left(  \hat{B}_{a}^{\dag}(t)\hat{B}%
_{b}(t-s)\rho^{I}_{\mathrm{B}}\right) \nonumber \\
&  -\sum_{a,b}\sum_{\omega,\omega_{1}} e^{i\left(  \omega
_{1}-\omega \right)  t}\hat{A}_{a}(\omega)\hat{A}_{b}(\omega)\rho
^{I}_{\mathrm{S}}(t)\int_{0}^{t}dse^{i\omega s}\mathrm{tr}_{\mathrm{B}}\left(  \hat
{B}_{a}^{\dag}(t)\hat{B}_{b}(t-s)\rho^{I}_{\mathrm{B}}\right)    + H.c.,
\end{align}
\end{widetext}
where $a,b=1,2,3,...,L $. We employ the reservoir correlation functions%
\begin{align}
\Gamma_{ab}\left(  \omega \right)   &  =\int_{0}^{t}dse^{i\omega s}%
\mathrm{tr}_{\mathrm{B}}\left(  \hat{B}_{a}^{\dag}(t)\hat{B}_{b}(t-s)\rho
^{I}_{\mathrm{B}}\right)   
\end{align}
to simplify the above equation. Eventually, $\frac{d}{dt}\rho_{I}^{\mathrm{S}%
}(t)$ of the system S in the interaction picture is
\begin{align}
\frac{d}{dt}\rho^{I}_{\mathrm{S}}(t)  &  =\sum_{a,b}\sum_{\omega,\omega_{1}}
e^{i\left(  \omega_{1}-\omega \right)  t}\Gamma_{ab}\left(  \omega \right)  [
\hat{A}_{b}(\omega)\rho^{I}_{\mathrm{S}}(t)A_{a}^{\dag}(\omega_{1})\\
&  -\hat{A}_{a}(-\omega_{1})\hat{A}_{b}(\omega)\rho^{I}_{\mathrm{S}}(t) ]
+H.c.\nonumber
\end{align}
We use the rotating wave approximation to average out the high-frequency part
of quantum transition processes and ignore the case of $\omega \neq \omega_{1}$,
then get%
\begin{align}
\frac{d}{dt}\rho^{I}_{\mathrm{S}}(t)  &  =\sum_{a,b}\sum_{\omega} \{ \Gamma
_{ab}\left(  \omega \right)  [\hat{A}_{b}(\omega)\rho^{I}_{\mathrm{S}}%
(t)\hat{A}_{a}^{\dag}(\omega)\nonumber  \\
&  -\hat{A}_{a}(-\omega)\hat{A}_{b}(\omega)\rho^{I}_{\mathrm{S}}(t)]+H.c.\}. \label{52}
\end{align}

\bigskip

\bigskip

\bigskip

\section*{ Appendix B: Instability of  quasi-Hermitian system without conserved quantity}
\setcounter{equation}{0}
\renewcommand\theequation{B\arabic{equation}}
In this section, we demonstrate the instability of quasi-Hermitian systems by examining the diagonal elements of the steady-state density matrix. The diagonal elements of the density matrix represent the occupation probabilities of the corresponding states. By studying these elements, we can gain insights into the stability and dynamics of the quasi-Hermitian system.
 
By using Eq. (\ref{52}), the diagonal terms of density matrix defined as
$P(n,t)=\left \langle n\right \vert _{L}\rho_{\mathrm{S}}^{I}(t)\left \vert
n\right \rangle _{L}$ satisfy
\begin{equation}
\frac{d}{dt}P(n,t)=\sum_{m}[W(n|m)P(m,t)-T(m|n)P(n,t)], \label{13}%
\end{equation}
where%

\begin{equation*}
W(n|m)=\sum_{a,b}\gamma_{ab}(E_{m}-E_{n})\left \langle m\right \vert _{R}%
\gamma_{a}\hat{C}_{a}\left \vert n\right \rangle _{L}\left \langle n\right \vert
_{L}\gamma_{b}\hat{C}_{b}^{\dag}\left \vert m\right \rangle _{R},
\end{equation*}
\begin{equation}
T(n|m)=\sum_{a,b}\gamma_{ab}(E_{m}-E_{n})Re(F_{ab,mn})+S_{ab}Im(F_{ab,mn}),
\end{equation}
here $\left \vert n\right \rangle _{L}$ is the  left-eigenstate which corresponding to $\left \vert n\right \rangle _{R}$, and  satisfy $_{R} \left \langle n | n\right \rangle
_{L}=1$. $F_{ab,mn}=\gamma_{a}\gamma_{b}\left \langle m\right \vert _{R}\hat{C}%
_{a}\left \vert n\right \rangle _{L}\left \langle n\right \vert _{R}\hat{C}%
_{b}^{\dag}\left \vert m\right \rangle _{L}$, $ \gamma_{ab}$ is the real part of 2$\Gamma_{ab}$, 
and $S_{ab}$ is the
image part of 2$\Gamma_{ab}.$

 We define a matrix $A$, which element is $A_{nn}%
=W(n|n)-\sum_{m}T(m|n)$, and $A_{mn}=W(m|n).$ 
A physical steady state solution is exist, it must be required
$\det(A)=0$. Otherwise, it means that there is no steady state in the system.
However, generally speaking, when there is no special constraint, the
determinant of a matrix will not be 0. This means that in general,
quasi-Hermitian systems have no steady state. In order to make the system have a
steady state, and the steady state of the system has a certain stability, for
example, it will not be destroyed by the slight change of Hamiltonian in the
thermal bath. The steady state of the system is determined by the
relative proportion of each matrix element. When adjusting the Hamiltonian of
the thermal bath, we can change the relative size of the real part and
imaginary part of the system and affect the proportion of matrix elements. Therefore, a
necessary requirement to keep the system stability is $ImF_{aa,mn}=0$. for all $a$ and $m,n$. This means 
\[
\left \langle m\right \vert _{R}\hat{C}_{a}\left \vert n\right \rangle
_{L} \propto \left \langle n\right \vert _{R}\hat{C}_{a}^{\dag}\left \vert
m\right \rangle _{L}^{\ast}%
\]
or%
\[
\left \langle m\right \vert _{0}\hat{\mathcal{S}}^{\dag}\hat{C}_{a}\hat{\mathcal{S}}^{-1}\left \vert
n\right \rangle _{0}=k\left \langle n\right \vert _{0}\hat{\mathcal{S}}^{\dag}\hat{C}_{a}^{\dag
}\hat{\mathcal{S}}^{-1}\left \vert m\right \rangle _{0}^{\ast}%
\]
\[
\left \langle m\right \vert _{0}\hat{\mathcal{S}}^{\dag}\hat{C}_{a}\hat{\mathcal{S}}^{-1}\left \vert n\right \rangle _{0}=k\left \langle n\right \vert _{0}\hat{\mathcal{S}}^{\dag}\hat{C}_{a}^{\dag
}\hat{\mathcal{S}}^{-1}\left \vert m\right \rangle _{0}^{\ast}
\]
or
\begin{equation}
\mathcal{T}_c\hat{C}_{a}\ =k\hat{C}_{a}\mathcal{T}_c.
\end{equation}

Take trance on both sides of the equation, we get%
\begin{equation}
\mathrm{tr}(\mathcal{T}_c\hat{C}_{a})\ =k \mathrm{tr}(\mathcal{T}_c\hat{C}_{a}).
\end{equation}
This is equivalent to $k=1$ or $\mathrm{tr}(\mathcal{T}_c\hat{C}_{a})=0$. We think that the small change of Hamiltonian of thermal bath should not
affect the steady state of the system. Therefore, we can add a constant to $\hat{C}_{a}$
so that $\mathrm{tr}(\mathcal{T}_c\hat{C}_{a})\neq 0$. Therefore, we get $k=1$.  We have
\begin{equation}
\lbrack \mathcal{T}_c,\hat{C}_{a}]\ =0.
\end{equation}

In other words, the necessary condition for a system to have a stable state is
that the system has a conserved quantity. Conversely, it can be proved that
$\det(A)=0$ if the system has a conserved quantity. Recalled the definition
\begin{align}
\gamma_{ab} &  =\int_{-\infty}^{\infty}dte^{i\omega t}tr_\mathrm{B}\left(  \rho
_\mathrm{B}\hat{B}_{a}^{\dag}(t)\hat{B}_{b}(0)\right)  \nonumber \\
&  \equiv \int_{-\infty}^{\infty}dte^{i\omega t}\left \langle \hat{B}_{a}^{\dag
}(t)\hat{B}_{b}(0)\right \rangle,
\end{align}
and using the Kubo-Martin-Schwinger (KMS) condition $\left \langle \hat{B}%
_{a}^{\dag}(t)\hat{B}_{b}(0)\right \rangle =\left \langle \hat{B}_{b}(0)\hat
{B}_{a}^{\dag}(t+i\frac{1}{T})\right \rangle $, we derive the temperature
dependent behavior of $\gamma_{ab},$ i.e.,
\begin{equation}
\gamma_{ab}(-\omega)=e^{-\omega/T}\gamma_{ba}(\omega).\label{14}%
\end{equation}

We  can deduce that  $T(n|m)e_{{}}^{-E_{n}/T}W_{n}=\ W(m|n)e^{-E_{m}%
/T}W_{m}$. It can be verified $\det(A)=0$ and the steady-state solution
given by $P_{n}\propto W_{n}e^{-\beta_{T}E_{n}}$ which is consistent with the
result obtained in the  text.
\bigskip

\section*{ Appendix C: Uniqueness of $\mathcal{T}_c$ for thermalizable quasi-Hermitian systems}
\setcounter{equation}{0}
\renewcommand\theequation{C\arabic{equation}}

In this appendix, we aim to reach a consensus that a thermalizable non-Hermitian system can be characterized by a unique  $\mathcal{T}_c$ (up to an irrelevant multiplicative factor that leaves the expectation values of physical quantities unaffected).

To establish this, we first consider the restrictions of ergodicity of the non-Hermitian system under investigation on the operator $\mathcal{T}_c$. Consider a subsystem. Ergodicity can be interpreted as follows: if the initial state of the subsystem is confined within a particular subspace, and continues to remain within that same subspace after time evolution, it indicates that the system lacks ergodicity.  Conversely, if the system possesses ergodicity, then there does not exist such a subspace. 

Assume two distinct $\mathcal{T}_c$ satisfying Lemma 1, represented as $\mathcal{T}_{c1}$ and $\mathcal{T}_{c2}$. According to Lemma 1's first condition, they can be expressed as:
\begin{equation}
\mathcal{T}_{ci}=\sum_{m}\  t_{mi} \left \vert m\right \rangle _{R}\left \langle
m\right \vert _{R},
\end{equation}
where $i=1,2$, $\left \vert m\right \rangle _{R}/\left \vert m\right \rangle _{L}$ represent the biorthogonality right/left eigenstates of $\hat{H}_{\mathrm{NH}}$. $\{ t_{mi} \}$ is   a set of positive real number.
Now, define the operator 
\begin{equation}
\mathcal{T}_{*}=  \mathcal{T}_{c1}- \kappa \mathcal{T}_{c2},
\end{equation}
where $\kappa=min (\frac{t_{m1}}{t_{m2}})$.  It can be verified that:
\begin{equation}
 \mathcal{T}_{*} \left \vert m\right \rangle _{L} = t_{m2}(\frac{t_{m1}}{t_{m2}}-\kappa)\left \vert m\right \rangle _{R}.
\end{equation}
For cases where $\mathcal{T}_{c1}$ and $\mathcal{T}_{c2}$ are not merely different by a multiplicative factor, we can delineate a subspace:  \begin{equation}V^* = \left \{\left \vert v \right \rangle    \bigg| \left \vert v \right \rangle  = \sum_{k, \frac{t_{k1}}{t_{k2}}=\kappa
}  \lambda_k \left|k\right\rangle_L ,  \lambda_k \in \mathbb{C} \right \} .\end{equation}  
 As preparation for further discussion,  we   calculation $\mathrm{tr} (\mathcal{T}_{*} \rho )$ for a arbitrary ensemble $\rho=\sum_{i}  P_i \left \vert i\right \rangle  \left \langle
i\right \vert  $. We have
\begin{align}
\mathrm{tr} (\mathcal{T}_{*} \rho )&  =  \sum_i P_i \langle i | \mathcal{T}_{*} | i \rangle \notag  \\
& = \sum_i \sum_m P_i |\left  \langle i | m\right \rangle _{R}|^2  t_{m2}(\frac{t_{m1}}{t_{m2}}-\kappa) .
\end{align}

Based on the above expression, if a quantum state exists in the ensemble that does not belong to the subspace $V^*$, we have $\mathrm{tr} (\mathcal{T}_{*} \rho ) \neq 0$. Conversely, if all states belong to the subspace V, we have $\mathrm{tr} (\mathcal{T}_{*} \rho ) = 0$.

Suppose the system's ensemble at $t=0$ mixes states within the subspace $V^*$, i.e., $\rho=\sum_{i} P_i \left \vert v_i\right \rangle \left \langle v_i\right \vert $, where $\left \vert v_i\right \rangle \in V^*$. We can verify that $\mathcal{T}_{*}$ also adheres to conditions (1) and (2) of Lemma 1. Hence, $P_c(t)\mathcal{T}_{*}$ is a conserved quantity. The existence of the conserved quantity $P_c(t)\mathcal{T}_{*}$ yields:
\begin{equation}
\mathrm{tr} (\mathcal{T}_{*} \rho(t)) = \mathrm{tr} (\mathcal{T}_{*} \rho(t=0)) / P_c(t) =0.
\end{equation}
This fact signifies that if the states within the ensemble originated in the subspace $V^*$ at the initial time, post the time evolution, the states within the ensemble will continue to reside in the subspace $V^*$. In such a scenario, $P_c(t)\mathcal{T}_{*}$ acting as a path-independent conserved quantity, causes the break of the system's ergodicity. 
Thus, it is observed that ergodic non-Hermitian systems possess at most one conserved quantity.

Notably, both non-Hermitian and Hermitian systems, interconnected by the similarity transformation $\hat{\mathcal{S}}$, share ergodicity. If the non-Hermitian system possesses an subspace $V^*$ that breaks  ergodicity, then the corresponding Hermitian system will have an  subspace $V^*_0 = \left \{\hat{\mathcal{S}}^{-1}\left \vert v \right \rangle \big| \left \vert v \right \rangle \in V^* \right \}$ that also breaks ergodicity. Because the thermalizable Hermitian system exhibits ergodicity, therefore, the thermalizable non-Hermitian system exhibits ergodicity as well. This indicates that up to an irrelevant multiplicative factor, a thermalizable non-Hermitian system has a unique $\mathcal{T}_c$.

\begin{figure*}[ptb]
\includegraphics[width=12cm]{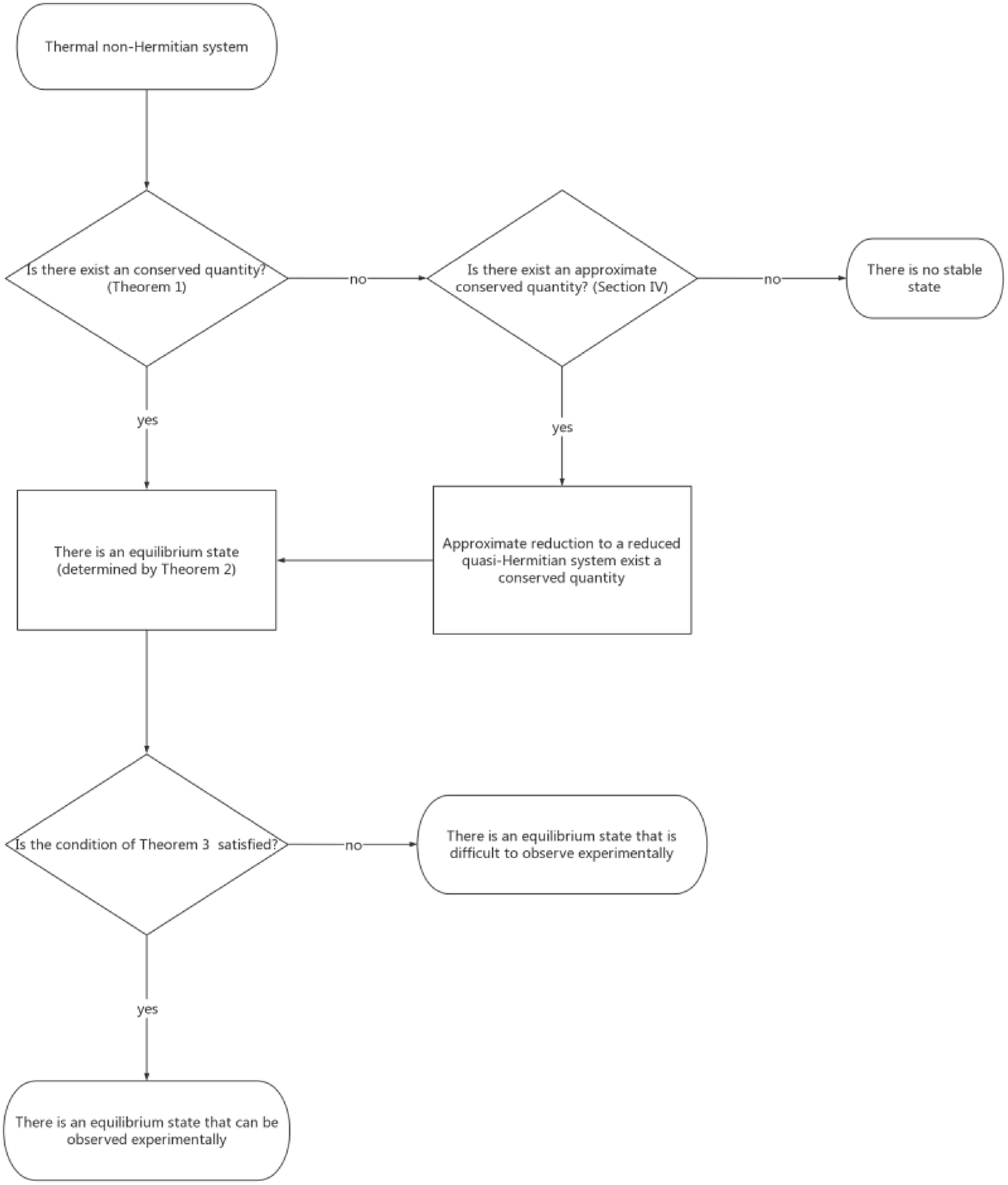}
\caption{The algorithm for determining the equilibrium state of non-Hermitian system.}
\end{figure*}

\section*{ Appendix D: Reduce  Non-Hermitian System to Quasi-Hermitian System}
\setcounter{equation}{0}
\renewcommand\theequation{D\arabic{equation}}
First, we present the master equation of a general non-Hermitian system. We consider the evolution equation of a non-Hermitian system under such an interaction representation:
\begin{align}
\frac{d}{dt}\rho^{I}_{\mathrm{S+B}}(t)=&-i[\hat{V}_{I}\left(  t\right)
\rho^{I}_{\mathrm{S+B}}(t)-\rho^{I}_{\mathrm{S+B}}(t)\hat{V}_{I}^{\dag}\left(
t\right)  ] \notag \\
&+ (Im \hat{H}_{\mathrm{eff},0} 
\rho^{I}_{\mathrm{S+B}}(t)+\rho^{I}_{\mathrm{S+B}}(t) Im \hat{H}^\dag_{\mathrm{eff},0} )  ,  
\end{align}
where $\rho^{I}_{\mathrm{S+B}}(t)=e^{i Re\hat{H}_{\mathrm{eff},0}t}%
\rho_{\mathrm{S+B}}(t)e^{-i Re\hat{H}_{\mathrm{eff},0}^{\dag}t}$ and $\hat
{V}_{I}\left(  t\right)  =e^{i Re \hat{H}_{\mathrm{eff},0}t}\hat
{H}_{BS}e^{-i Re \hat{H}_{\mathrm{eff},0}t}  $ is the interaction Hamiltonian in the interaction picture
and $\hat{H}_{\mathrm{eff},0}=\hat{H}_{\mathrm{NH}}\otimes \hat{I}_{ B}+\hat{I}_{ S }\otimes \hat{H}_{ B }$. The operator $Re \hat{H}_{\mathrm{eff},0}$ is defined as sharing the same eigenstates with $\hat{H}_{\mathrm{eff},0}$, but with eigenvalues corresponding to the real part of $\hat{H}_{\mathrm{eff},0}$'s eigenvalues. Similarly, the operator $Im \hat{H}_{\mathrm{eff},0}$ is defined as sharing the same eigenstates with $\hat{H}_{\mathrm{eff},0}$, but with eigenvalues corresponding to the imaginary part of $\hat{H}_{\mathrm{eff},0}$'s eigenvalues. Both operators are quasi-Hermitian. Compared with the
evolution equation of a quasi-Hermitian system $Re \hat{H}_{\mathrm{eff},0}$, the evolution equation of
a non-Hermitian system has one more term. Let's explain how this term which representing dissipation 
reduces the steady state of the system to that of the reduced quasi-Hermitian system.

We note that dissipation will mainly introduce a new term to the Liouville
superoperator:
\begin{equation}
\mathcal{L}_{diss \ mn,kl}= (\lambda_{m}+\lambda_{n})\delta_{mn,kl}.
\end{equation}
Here, $\lambda_{n}$ is the imaginary part of the eigenvalue corresponding to the eigenstate labeled by $n$. Under the condition of strong dissipation, the Liouville superoperator
introduced by the thermal bath's coupling should be treated as a perturbation
term. According to degenerate perturbation theory,
in leading order, the state with the maximum eigenvalue of the original Liouville super operator $\mathcal{L}_{\mathrm{NH}}$ is
equivalent to the state with maximum eigenvalue of $\mathcal{L}^R$ which
satisfies:
\begin{equation}
\mathcal{L}^R= \mathcal{P} \mathcal{L}_{\mathrm{NH}} \mathcal{P},
\end{equation}
where $\mathcal{P}$ is the projection super operator projecting to the subspace with
$\mathcal{L}_{diss} \rho=0$. That is, $ \mathcal{P}(\rho)=\rho$ for $\mathcal{L}_{diss} \rho=0$, and $\mathcal{P}(\rho)=0$ for $\mathcal{L}_{diss} \rho \neq 0$. This is equivalent to solving the quasi-Hermitian
problem with:
\begin{equation}
\hat{H}^R_{\mathrm{NH}} = P\hat{H}_{\mathrm{NH}}P,
\end{equation}
and coupling:
\begin{equation}
\hat{H}_{BS}^{R}=\sum_{a}\lambda_a [( P \hat{C}_{a} P )\otimes \hat{B}_{a}+( P \hat{C}^\dag_{a} P)\otimes \hat{B}^\dag_{a}]  .
\end{equation}
Here, $P$ is defined by $P \hat{O} P=\mathcal{P} (\hat{O} )$.

To analyze the applicability of this approximation, we examine the corrections to the steady-state density matrix due to transitions between non-dissipative and dissipative states. Within the first-order range, the presence of this transition term does not impact the density matrix of the system within the subspace of non-dissipative states. We only need to analyze the portion of the density matrix that does not belong to the space of dissipative states.

Firstly, we consider the off-diagonal matrix elements of the density matrix. This non-Hermitian term contributes an additional decoherence rate, which causes the off-diagonal terms outside the subspace to vanish rapidly. Therefore, the off-diagonal terms are all zero. Secondly, we examine the diagonal matrix elements of the density matrix. In the non-Hermitian case, when $\vert \lambda_{n}\vert \gg \vert  \gamma_{a} \vert $, the solution can be derived using perturbation theory. To this order, we obtain the probability of the system being in the dissipative states $n$ as $P_{n}^{(1)} = -\sum_s  \frac{W(n|s)}{ 2\lambda_{n}  } P_{s} \sim \frac{\tau_{diss}}{\tau_{nondiss}}$. Here, $ P_{s} $ represents the probability of the steady-state density matrix's eigenstates in the non-dissipative states. $\tau_{diss}$ represents the dissipative time scale, and $\tau_{nondiss}$ denotes the time scale of evolution for non-dissipative states. Under the weak coupling limit, this term is indeed approximately zero. Therefore, we show the effectiveness of the reduction method considered in the text under the weak coupling limit.

\section*{ Appendix E: Algorithm for determining the equilibrium state of non-Hermitian system}

 In this appendix, we provide an algorithm (Fig. 4) for determining the equilibrium state of non-Hermitian systems as a useful summary of the paper. To avoid excessive complexity in the algorithm, we assume that all quasi-Hermitian systems without quasi-Hermiticity breaking are thermalizable quasi-Hermitian systems. While this assumption is not strictly accurate, it holds true for the vast majority of cases.

\end{document}